\documentclass[acmsmall]{acmart}
\AtBeginDocument{%
  \providecommand\BibTeX{{%
    \normalfont B\kern-0.5em{\scshape i\kern-0.25em b}\kern-0.8em\TeX}}}

\setcopyright{rightsretained}
\acmJournal{PACMMOD}
\acmYear{2024} 
\acmVolume{2} 
\acmNumber{1 (SIGMOD)} 
\acmArticle{21} 
\acmMonth{2} 
\acmDOI{10.1145/3639276}

\usepackage{xspace,xpunctuate}
\newcommand{\ie}{{i.e.,}\xspace}

\newcommand{\eg}{{e.g.,}\xspace}
\newcommand{\ea}{{et~al\xperiod}\xspace}

\newcommand{\etc}{{etc\xperiod}\xspace}

\newcommand{\revise}[1]{{\color{black}#1}}

\usepackage{multirow}
\usepackage{booktabs}

\received{July 2023}
\received[revised]{October 2023}
\received[accepted]{November 2023}
\begin{document}

\title{Optimizing Dataflow Systems for Scalable Interactive Visualization}

\author{Junran Yang}
\email{junran@cs.washington.edu}
\orcid{0000-0002-8467-2917}
\affiliation{%
  \institution{University of Washington}
  \city{Seattle}
  \country{USA}
}

\author{Hyekang Kevin Joo}
\affiliation{%
  \institution{Carnegie Mellon University}
  \orcid{0000-0002-6387-5686}
  \city{Pittsburgh}
  \country{USA}}
\email{kevinjoo@andrew.cmu.edu}

\author{Sai Yerramreddy}
\orcid{0000-0003-0848-6351}
\affiliation{%
  \institution{University of Maryland}
  \city{Collkge Park}
  \country{USA}
}
\email{saiyr@umd.edu}

\author{Dominik Moritz}
\affiliation{%
  \institution{Carnegie Mellon University}
  \orcid{0000-0002-3110-1053}
  \city{Pittsburgh}
  \country{USA}}
\email{domoritz@cmu.edu}

\author{Leilani Battle}
\orcid{0000-0003-3870-636X}
\email{leibatt@cs.washington.edu}
\affiliation{%
  \institution{University of Washington}
  \city{Seattle}
  \country{USA}
}


 \begin{abstract}


Supporting the interactive exploration of large datasets is a popular and challenging use case for data management systems.
Traditionally, the interface and the back-end system are built and optimized separately, and interface design and system optimization require different skill sets that are difficult for one person to master. 
To enable analysts to focus on visualization design, we contribute VegaPlus, a system that automatically optimizes interactive dashboards to support large datasets.
To achieve this, VegaPlus leverages two core ideas. First, we introduce an optimizer that can reason about execution plans in Vega, a back-end DBMS, or a mix of both environments. The optimizer also considers how user interactions may alter execution plan performance, and can partially or fully rewrite the plans when needed.
Through a series of benchmark experiments on seven different dashboard designs, our results show that VegaPlus provides superior performance and versatility compared to standard dashboard optimization techniques.

\end{abstract}



\maketitle

\section{introduction}
\label{sec:introduction}

Interactive data visualization is essential for understanding, manipulating and presenting complex datasets~\cite{tufte1985visual, card1999readings}. Specification languages such as D3~\cite{bostock_d_2011}, Plotly~\cite{plotly} and Vega~\cite{satyanarayan_reactive_2016}
make the process of designing interactive visualizations more systematic, precise and accessible to both novices and analysts. Meanwhile, they 
are also the building blocks of all visualization systems. For example, Forcache~\cite{battle_dynamic_2016} and Kyrix~\cite{tao_kyrix_2019-1} are implemented using D3~\cite{bostock_d_2011} while Falcon~\cite{moritz_falcon_2019} and Voyager~\cite{2017-voyager2} are built on top of Vega/Vega-Lite~\cite{satyanarayan_vega-lite_2017, satyanarayan_reactive_2016}. Hence, enhancing visualization languages can not only extend their usability but also improve all of the systems they support, leading to a cumulative impact on the data visualization ecosystem. 
\revise{However, these languages are not designed to optimize data queries, and, as a consequence, they fail to match the data processing performance of even the most basic database systems~\cite{moritz2015dynamic}.}

\revise{A natural next question is: why not simply add classic data management optimizations and/or cost models to visualization languages? Prior works show how naive applications of classic database optimization methods fail to support interactive visualization scenarios~\cite{battle2020database,battle_dynamic_2016,psallidas_smoke_2018,galakatos_revisiting_2017,liu2013immens}. In contrast, visualization- and interaction-aware optimization approaches are more effective for scaling up visualization systems~\cite{battle2020structured}. Moreover, with the huge diversity in possible visualization and dashboard designs, system configurations, and user interaction behaviors, it is difficult to hand-tune classic cost models for selecting efficient optimization plans in each analysis scenario.
While many query optimization} techniques have also been proposed to support specific interface or interaction designs~\cite{battle2020structured}, they often fail to translate to new visualization interfaces.
\revise{This makes developing an optimizer a major challenge for big data visualization and exploration systems.}
%
\revise{Furthermore, g}iven that our target users are not always system builders or DBMS administrators, 
\emph{an ideal solution is to build \revise{dynamic} optimizations directly into visualization languages}, so that the users can avoid bespoke optimization techniques.
\revise{In this paper, we explore the design space of possible optimization strategies for Vega~\cite{satyanarayan_reactive_2016}, a popular visualization language. To do this, we first extend Vega to coordinate execution with a back-end DBMS such as PostgreSQL~\cite{postgres} or DuckDB~\cite{raasveldt_duckdb_2019}. Then, we explore the space of possible optimization strategies by training machine learning models to predict efficient Vega+DBMS execution plans among hundreds of candidates derived from a variety of dashboard configurations. By inspecting what these models learn, we contribute robust heuristics towards the development of effective cost models for Vega. Based on our findings, we introduce \textsc{VegaPlus}, a system that automatically optimizes dashboards implemented in Vega by (a) rewriting Vega's data transformations as corresponding SQL queries and (b) selecting efficient plans for executing these queries across the client-side Vega runtime and server-side DBMS.}


A declarative Vega specification often contains a data transformation pipeline, \revise{i.e., a} visualization query that generates a \emph{dataflow graph} as the execution plan.
\revise{We contribute a \textbf{new query rewriting approach for Vega} that augments} selected dataflow operators
to emit a SQL query
that performs equivalent data manipulation operations (\eg \texttt{filtering}, \texttt{binning}, and \texttt{aggregation}).
In this way,
we can augment Vega to leverage the computational strengths of relational DBMSs like PostgreSQL~\cite{postgres} and DuckDB~\cite{raasveldt_duckdb_2019} with minimal changes to the user's visualization workflow.

However, naively pushing all computation to the DBMS can inadvertently \emph{increase} system latency, since it completely ignores available computing resources on the client and may introduce unnecessary round trips across the network. Furthermore, the DBMS is unaware of how user interactions may alter query execution across Vega dataflows, hindering its ability to optimize these dataflows on its own. For example, interactive filters (e.g., sliders, brushes) may or may not generate filter predicates on emitted queries, depending on the user's interaction choices.
\revise{In response, we explore a range of techniques for selecting efficient query execution plans in an interaction-aware way. Specifically, we contribute \textbf{a collection of machine learning models to predict efficient plans} given metadata about} dataflow structure, dataset characteristics, and anticipated execution costs for static \emph{and} interactive visualization scenarios. \revise{Then, we analyze these models to contribute \textbf{general-purpose heuristics} that can be used to develop robust cost models for Vega, which we refer to as \emph{heuristic-based models}.}
\revise{Furthermore, we extend these optimizations to}
exploit repetition in user interaction behaviors by checking when query plans may benefit from previously cached results, inspired by prior work~\cite{battle_dynamic_2016,terlecki_improving_2015}.

\revise{To identify the best models and heuristics for selecting efficient execution plans, we contribute \textbf{a Vega benchmark suite} that simulates interactions with seven different Vega dashboard templates. We use this benchmark to test the efficacy of our plan selection models across different visualization designs, input datasets, and interactive scenarios, inspired by prior work on visualization benchmarking}~\cite{eichmann_idebench_2020,battle_database_2020}.


\revise{\textbf{Our techniques can also generalize beyond Vega}} to any visualization language that compiles into dataflow graphs. Given a dataflow graph from another language, our query rewriting methods can easily be extended to map detected data transformation nodes to corresponding SQL queries, which can then be passed directly to the VegaPlus optimizer. These techniques can be particularly useful in development environments that contain significant information about the client but limited server-side control, e.g., in online notebook environments such as Observable~\cite{observable}.


In this paper, we make the following contributions:
\begin{itemize}
    \item Automated query rewriting for the Vega visualization language, which translates data transformations \revise{from Vega specifications} into SQL queries to be executed on a connected DBMS.
    \item \revise{A systematic analysis of Vega's query optimization space, using machine learning models to navigate hundreds of viable execution plans} that balance data processing between the client dataflow in Vega and an external DBMS.
    \item \revise{Comparison of multiple optimizer designs for supporting interactive visualization of large datasets, based on the results of our aforementioned analysis.}
    \item \revise{A new Vega performance benchmark and e}valuations demonstrating how \revise{the proposed} optimizer\revise{s scale up and speed} up visualization processing for a wide range of existing designs compared to Vega. 

\end{itemize}




\section{Background}
\label{sec:background}
VegaPlus aims to scale up interactive visualizations generated with Vega for large data. Here, we provide a brief overview of Vega and its dataflow system. 

\begin{figure}
    \includegraphics[width=.5\linewidth]{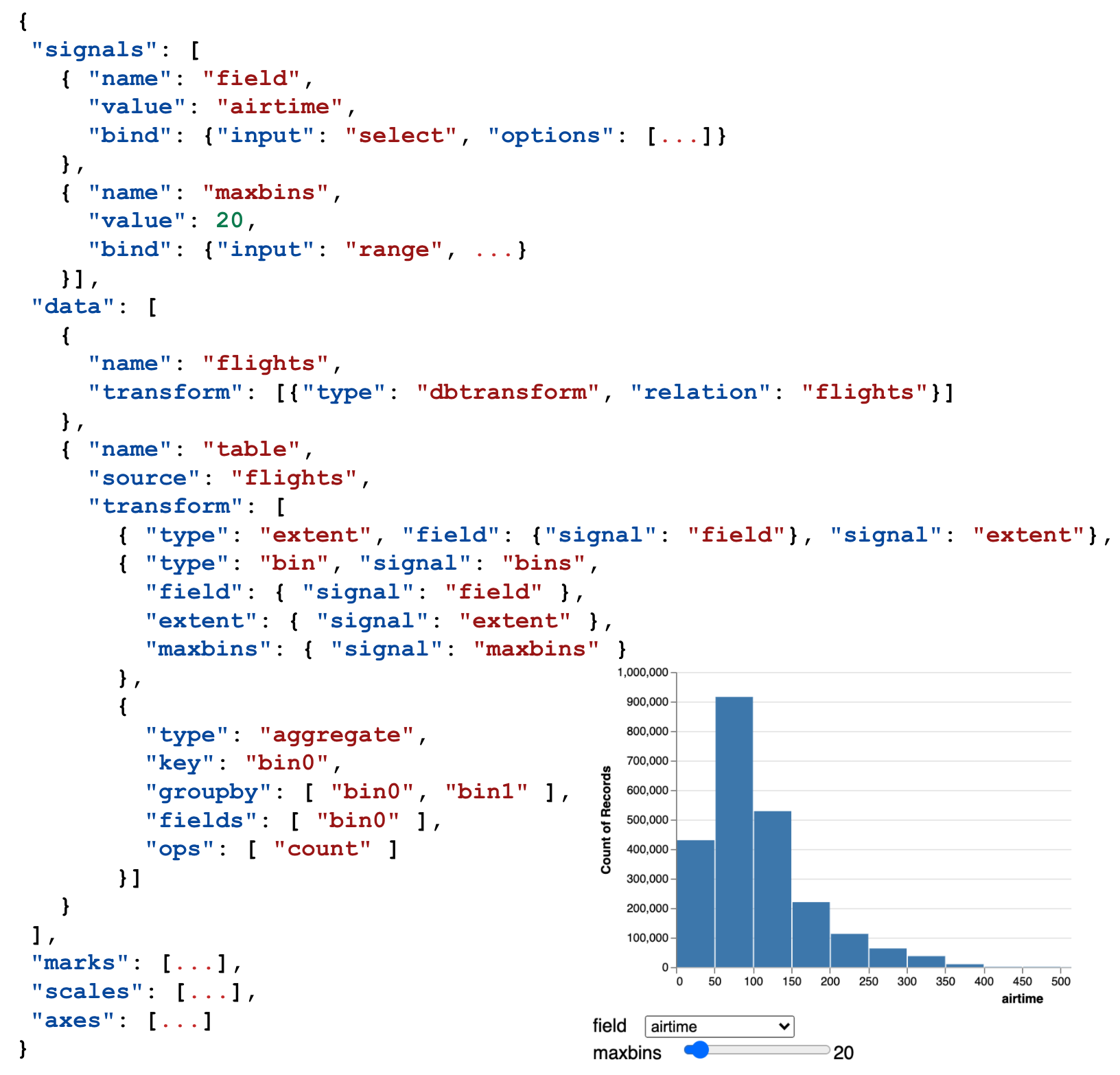}
     \caption{A histogram example with dynamic queries. Users can select a data field from the drop-down menu, or adjust the bin size with the slider. }
    \label{fig:flights}
\end{figure}


\paragraph{Vega Dataflow.}
The Vega runtime parses specifications in JSON to generate interactive, web-based graphics.
To efficiently process and route data to relevant visualization components and layers, Vega adopts a dataflow compilation and execution model~\cite{satyanarayan_reactive_2016}.
Dataflow systems are of particular interest when performing interactive, incremental stream processing in a distributed environment~\cite{gjengset_noria_2018,murray_incremental_2016,murray_naiad_2013,chandrasekaran_telegraphcq_2003}.
It is also a common data model in visualization systems (\eg Vega~\cite{satyanarayan_reactive_2016}, VTK~\cite{schroeder_design_1996}) where its operators form a directed graph.

As a declarative language, Vega decouples specification from execution, allowing customization and optimization for data processing in its dataflow system.
Given a user's declarative specification, Vega automatically constructs the corresponding dataflow graph.
When executed, the dataflow graph computes a series of data transformations through known operators (\eg filter, map, aggregate), where data records are processed by each operator as they pass through the graph. Finally, the computed data is mapped to visual encodings and positions. 
As an example, the Vega specification in \autoref{fig:flights} uses the \texttt{bin} and \texttt{aggregate} transforms to calculate the bins and counts for the rendered histogram.
Note that the instantiated dataflow graph at runtime may include additional internal operations that are not declared in the user's specification to support data processing, such as copying or sorting the data to facilitate the creation of axes, scales, etc.  

In VegaPlus, we propose new features to enable relational DBMSs to participate in the computation of Vega operators while preserving the original Vega dataflow structure. In this way, we enable users to connect visualization languages with data processing systems that they may not otherwise benefit from.

\paragraph{Data Pipeline. }
The \texttt{data} component (\ie data pipeline) in a specification is an array of \texttt{data entries}, and each data entry may contain a series of Vega \texttt{transforms} in the transform array, as in \autoref{fig:flights}. Each data entry points either to another data entry's output or the raw data as source. 
In the instantiated dataflow, corresponding operations maintain the same relative order as in the specification, even though internal operators are added. 
The data pipeline is transformed into the dataflow, which is a directed acyclic graph (DAG). \autoref{fig:rewrite} demonstrates a simplified version of the dataflow: the root nodes are signals and the data source references the DBMS tables, while non-root nodes are the transform operators.

\paragraph{From Specification to Dataflow Graph.}
Vega \texttt{Transforms} are the operators inside the dataflow graph for data processing and transformation. Each transform operator takes data tuples as input, performs certain computational operations on them, and then generates new data tuples as output. Vega includes a variety of common transform types that are particularly useful in constructing visualizations~\cite{satyanarayan_reactive_2016}.
Filtering, binning and aggregation are common examples of transform operations. 
Vega Transforms are typically specified within the transform array of a data definition. At compilation time, Vega instantiates a dataflow where the transform operators form a path in the order they are specified. Additional internal operators are added in between to calculate needed statistics or propagate updates. However, since the operators are evaluated in topological order, the way the data is processed is entirely decided by the specifications set by the user. Although SQL queries share the same declarative nature as Vega's visualization specifications, a DBMS typically rearranges and optimizes how the data are processed, producing a more efficient query execution structure overall. 
Hence, we propose techniques to partition the dataflow operations for execution within a DBMS while maintaining the relative orders.
However, we have also designed our dataflow rewriting methods to support extensions in the future.

\paragraph{Vega Parameters \& Signals.}
Parameters that specify an operator can either be fixed values or live references to other operators. Interaction events can update operator parameters or data inputs, and the changes are only re-evaluated by the operators downstream to the update. 
Further, users can track the interaction state in the Vega dataflow by specifying \texttt{signals}, i.e., special variables in the Vega specification used to capture the output of triggered interactions.
To this end, signals can be used to dynamically parameterize visual encoding properties or transform parameters via expression languages. For example, in \autoref{fig:flights}, two signals are specified: the first signal binds selections from the \texttt{field} dropdown menu to the \texttt{extent} transform in the specification and the second signal binds values from the \texttt{maxbins} slider to the number of bins calculated in the \texttt{bin} transform. Together, these signals control how the x-axis and bins (i.e., rectangles) are rendered in the histogram.
By extension, Vega signals become a form of data provenance that can be monitored by VegaPlus to detect changes in execution flow, e.g., to detect changes in the grouping/binning predicates in \autoref{fig:flights} when the user drags the slider, which would trigger recalculating the histogram bins.

\begin{example}\label{ex:background}
\textit{
Putting everything together, the example in Figure~\ref{fig:flights} shows a histogram visualization alongside its Vega specification. 
Its source dataset consists of flight arrival and departure details for all commercial flights in the USA from 1987 to 2008~\cite{flightDataset}. To explore the data distribution in terms of each data field, users can select the target data field from a drop-down menu and use the slider to find a desired binning granularity to summarize the record count. Although histograms are a common visualization type in data analysis, multiple operations are carried through when users interact with the chart. An \texttt{extent} transform is performed to calculate the x scale extent (minimum and maximum), which is also required by the \texttt{bin} transforms to calculate the binning buckets' start and end values. Other than the extent, the binning transform also needs the signals from both the \texttt{maxbins} slider and \texttt{field} drop-down menu. After the source data is distributed into discrete buckets, an \texttt{aggregate} transform counts the items inside each bucket. We use this example in the following sections to explain our optimization flow. 
}
\end{example}


\section{System Overview}
\label{sec:architecture}

\begin{figure}
    \includegraphics[width=.5\linewidth]{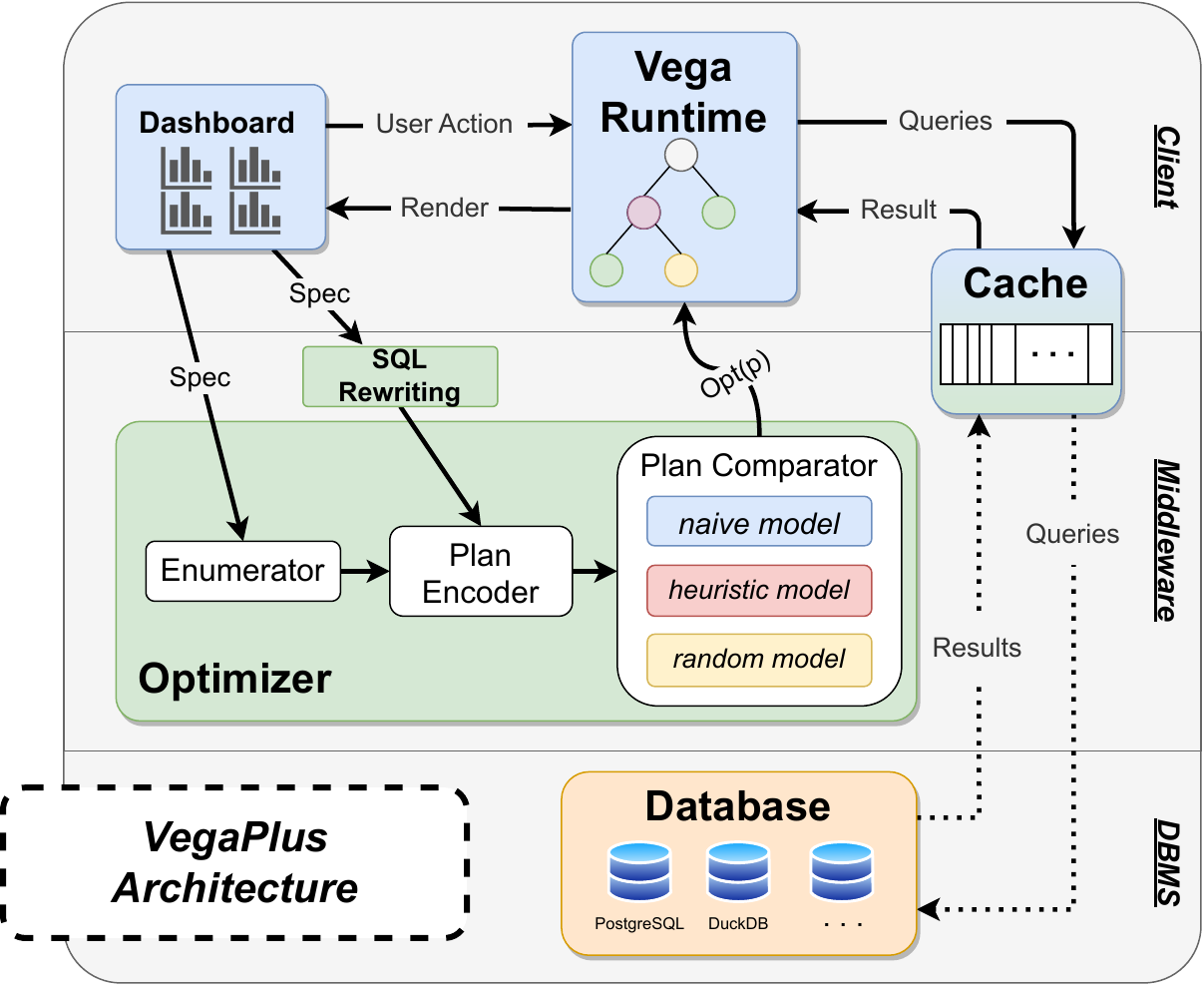}
     \caption{The architecture diagram. }
    \label{fig:arch}
\end{figure}

VegaPlus is comprised of three layers (see \autoref{fig:arch}): a client-side layer for rendering the visualizations, a server-side middleware layer for optimization, and a remote DBMS for scalable data processing. In this section, we summarize the functionality of each layer.

\paragraph{Client-Side Vega Runtime.} The Vega runtime generates the original Vega dataflow graph that renders the target visualization(s). The Vega runtime also manages any signals (i.e., interaction triggers) defined in the Vega specification, which VegaPlus needs to assess how interactions may alter the execution of the corresponding Vega dataflow graph. We have also augmented the Vega runtime to use specialized dataflow nodes that enable VegaPlus's SQL query rewriting functionality. Please see \autoref{sec:query-rewriting} for more details.

\paragraph{Server-Side Optimizer.} The optimizer is responsible for evaluating a given Vega dataflow graph, enumerating possible execution plans involving client and/or server resources, and selecting an efficient plan. To enumerate plans, the optimizer must coordinate with the client-side Vega runtime and (possibly remote) DBMS. To evaluate plans, the optimizer translates each plan into a vector encoding that can be recognized by machine learning models. Then, models are used to predict the most efficient plan. Finally, these predictions are used to dispatch queries either to the Vega runtime for local execution or to the DBMS for server-side or remote execution. We describe the optimizer further in \autoref{sec:optimization}.

\paragraph{DBMS.} VegaPlus gives users the option to share connection information for a DBMS within a Vega specification. This DBMS can be local, e.g., running on the client or even running in the browser, or it can be remote. VegaPlus uses this DBMS as a back-end for data processing. Specifically, any transforms from the Vega dataflow that are selected for SQL-based execution by the optimizer will be processed using the DBMS. To do this, VegaPlus issues the target queries to the DBMS using the specified connection, retrieves the results, then passes them back to the Vega runtime to be rendered. VegaPlus also uses the DBMS's plan analyzers to optimize Vega dataflow queries. For example, VegaPlus leverages the DBMS explain command to estimate execution costs. Please see \autoref{sec:optimization} for more details.

\section{Query Rewriting: From Vega Operators to SQL queries}
\label{sec:query-rewriting}

To facilitate interactive exploration in visualizations, transform operations are often one of the most time- and resource-intensive components. 
To address this problem, the query rewriter builds a bridge for VegaPlus to offload computationally-intensive Vega transforms as SQL queries executed on the backend DBMS. 
\revise{Specifically, we parse the Vega JSON specification and traverse the data pipeline to rewrite Vega transform operations into SQL queries. Then, we replace those rewritten operators with custom \texttt{VegaDBMSTransform} (VDT) operators. At runtime, when a VDT node is triggered, it builds and issues the corresponding SQL query and fetches the results via the middleware server. }


\paragraph{Candidate Transforms for Rewriting.} The VegaPlus query rewriter identifies applicable Vega transform nodes in the dataflow graph, such as
extent (\ie range values), bin, aggregate, filter, collect (\ie sorting), projection and stack (\ie window functions), automatically parses the transform parameters (if they exist), and translates the operation to a SQL query string builder using pre-defined templates implemented for each transform type. 
We targeted Vega transforms that match the select, project, group, and aggregate operations typically observed in OLAP queries to support offloading these operations to applicable DBMSs.
That being said, these SQL query builders often contain dynamic variables that are placeholders for signals or the output of parent nodes earlier in the dataflow.
Thus, VegaPlus must balance code and SQL query integration in a way that other visualization or database contexts may not.
For example, VegaPlus tracks how each node's dependencies are evaluated and propagated through the dataflow graph, allowing it to update the corresponding transform node with a complete SQL query when the holes in the query builder are filled. 
\revise{Translation of some common transform operators are straightforward due to the structured JSON format and clearly defined parameters such as the ``groupby'', ``ops'' and ``fields'' for \texttt{aggregate}. The \texttt{filter} transform is more complex to translate. It contains a predicate expression in a JavaScript-like expression language (\eg{\texttt{ \{"type": "filter", "expr": "datum.delay > 10 \&\& datum.delay < 30"\}}}), which evaluates to true or false to filter each data object. We parse the expression string into an Abstract Syntax Tree (AST) to generate a WHERE clause in SQL (\eg{ WHERE delay > 10 AND delay < 30}). Note that there is not always a one-to-one mapping between these Vega expressions and SQL. When an equivalent SQL predicate is not found, we fall back to native execution in Vega. }


\begin{figure}
    \includegraphics[width=.5\linewidth]{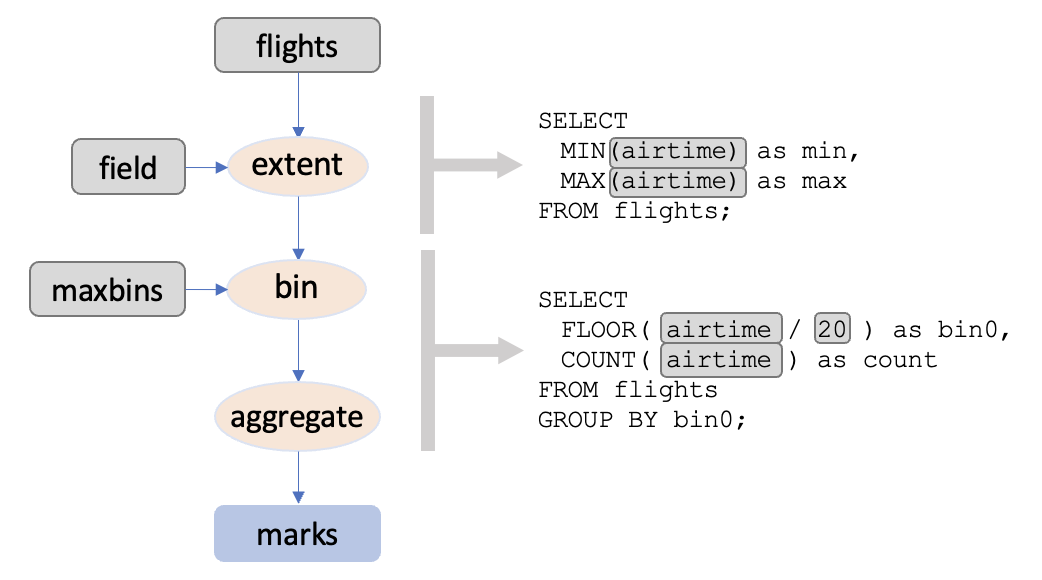}
     \caption{One possible way to rewrite the queries for the histogram example in Figure \ref{fig:flights}. The queries are parameterized by the input signals. The augmented nodes (VDTs) will re-evaluate and send queries to the DBMS when signals update. }
    \label{fig:rewrite}
\end{figure}


\paragraph{Efficient Transfers Between Vega and a DBMS} 
\revise{To automate two-way communication between the client and DBMS, VegaPlus replaces the pipelines of Vega transforms with \texttt{VegaDBMSTransform} (VDT),} a custom transform type that is responsible for \revise{building and} executing SQL queries against a DBMS.  
VDTs act as data sources in the data entry of a specification as in ~\autoref{fig:flights}.


VDTs are atypical Vega transforms in that they do not take any input data tuples from the upstream Vega dataflow; rather, the input data tuples can be thought of as the tuples from the \revise{DBMS (e.g., PostgreSQL)} in relation to the VDT targets. When a VDT's transform function is executed as part of dataflow graph execution in the browser, it sends an HTTP request containing an SQL query to a middleware server and waits for the results. The middleware server then forwards the query to a DBMS (PostgreSQL in this case) and then returns the query results to the browser. When the VDT in the browser receives these results, it emits them for propagation to downstream nodes in the Vega dataflow graph. It is important to note that VDT transform functions are blocking, synchronous operations. 
\revise{So far, we have assumed query rewriting based on individual transform operators. However, each VDT requires the query result to be returned to the client and stored in the dataflow graph, which leads to extra data transfers and memory consumption. To address this challenge, we support recursive rewriting of multiple transforms into a single nested query, i.e., batching transforms to reduce round trips. For example, an upstream transform can be rewritten as a subquery in downstream translation. Further, we also support rule-based query rewriting to transform nested batch queries into a more readable format, as shown in Example~\ref{ex:rewrite}. By default, our HTTP connector supports JSON, which requires client-side decoding and leads to large serialization overhead. To further reduce network transfer costs, VegaPlus encodes query results using the binary Apache Arrow format ~\cite{2023arrow}.}

\begin{example}\label{ex:rewrite}
\textit{In \autoref{fig:rewrite} we illustrate one way to rewrite the example specification in \autoref{fig:flights}. Suppose the VegaPlus optimizer emits an execution plan that rewrites the original \texttt{extent}, \texttt{bin}, and \texttt{aggregate} transforms using two VDTs: one for the \texttt{extent} and the other for \texttt{bin} and \texttt{aggregate}. The \texttt{extent} query is in a separate VDT because its output is referenced by the \texttt{scale} and the \texttt{bin} operator as a \texttt{signal}. At dataflow evaluation time, the \texttt{extent} VDT's evaluation function sends its SQL query to the middleware layer for execution. Once executed by the DBMS, the middleware sends the query results back to the Vega runtime, which get propagated through the dataflow graph
to the \texttt{bin-and-aggregate} VDT, where the bin's step size is calculated to complete the query string. Then, this query is evaluated in the same manner as the previous \texttt{extent} VDT.}
\end{example}

\section{Optimization}
\label{sec:optimization}

VegaPlus optimizes visualization dataflow execution by finding the most suitable plan to execute the corresponding Vega transforms and propagating the results to the Vega renderer. To achieve this, the optimizer enumerates all valid possibilities for partitioning the dataflow node execution across the client and server (\ie execution plans). Then, these plans are encoded into feature vectors to train models to predict the most suitable plan. In this section, we \revise{first explain the challenges of developing cost models for visualization (\autoref{sec:challenges}), and we then} describe the plan enumeration \revise{(\autoref{sec:enumeration})} and ranking strategies \revise{(\autoref{sec:comparison-model})} implemented in the VegaPlus optimizer.

\subsection{Visualization Optimization Challenges} \label{sec:challenges}

\revise{
There exist unique challenges for interactive visualizations.
First, with VegaPlus's cross-stack setup, the optimizer needs to coordinate between the backend user-provided DBMS, Vega runtime and the data transferring cost. 
\revise{For example, although MonetDB, DuckDB and PostgreSQL all have baseline cost models and query optimization methods that work well for their target use cases ~\cite{raasveldt_duckdb_2019, postgres}, they fail to provide adequate performance under real-time interactive visualization scenarios ~\cite{battle2020database}.}
Cost models are difficult to tune, especially when agnostic of the chosen systems. 
However, our goal is to efficiently utilize available resources with any provided configuration. 
Furthermore, a ready-to-use cost model for the dataflow operations in Vega currently does \underline{not} exist, making \revise{a fair} performance comparison nearly infeasible. Second, visualization workloads are different from typical OLAP workloads. A visualization can generate different queries \revise{depending on which interactions the user performs in the visualization interface}. Thus, the best plan that works for a static \revise{dashboard} might not perform well in the long run due to shifts in the queries executed in response to user interactions. 
}

\revise{More broadly, machine learning techiniques~\cite{marcus_plan_2019, Venkataraman_ernest_2016, liu_cardinality_2015, siddiqui_cost_2020, marcus_neo_2019} have been explored to replace core components in database query optimizers for efficiency and accuracy. } 
Traditionally, heuristic-based cost models are used to find the query plan with the minimum estimated cost. Cost estimations can be convenient to compare different plans of the same query. However, they cannot be directly used as an approximation of execution time without extensive tuning, and neither can they be taken out of the context to compare across different queries. 
Most learned optimizers aim to predict query plan latencies, but still suffer from high training costs and difficulties in model updating required by data or system configuration changes. 

That being said, there are specific opportunities in visualization that we do not observe in more general optimization scenarios. For example, many Vega visualizations tend to involve the same combinations of data transformations, even across different visualization types. This reduces the number and diversity of scenarios required to develop effective optimization models. Furthermore, this constrained visualization design space may suggest that complex models such as deep learning models may not be needed to achieve good performance.

\paragraph{VegaPlus' Pairwise Approach. }
We propose comparator models that focus on the relative performance difference between a pair of plans. 
More recent work~\cite{zhu2023lero} sidesteps the issues with \textit{pointwise} (\ie predicting cost/execution time) by instead applying pairwise ranking of plans to identify better and worse plans.  
\revise{Similarly, ~\cite{ding2019ai} uses pairwise comparison for index tuning. 
}
\revise{In VegaPlus, we formulate the problem of comparing the execution cost of two visualization partitioning plans as a classification task. While traditional pointwise learned cost models learn to minimize the prediction error for each plan, a pairwise model learns to minimize the error of comparison~\cite{zhu2023lero, ding2019ai}, which is intuitively more robust to the goal of plan selection in optimizers. }
With successive pairwise comparisons, we should arrive at a relatively efficient plan for a given Vega specification. 

\revise{
The pairwise approach provides benefits beyond a model for pairwise comparison. They produce insights to build rule-based heuristic models and cost models. The traditional ML algorithms are more transparent and easier to interpret than complex neural networks and deep learning models. For example, we can inspect the feature weights of a trained linear model to understand the feature importance. In this case, heuristics and priority can be extracted and translated to heuristics to build a rule-based optimizer, or as a starting point towards an efficient cost model that adapts to data input, dashboard designs, user interactions, and system configurations. 
In ~\autoref{sec:comparison-model}, we elaborate on how we can synthesize baseline cost models as a by-product of training a comparison model using the linear RankSVM algorithm. We also explain our approach to mapping observations of feature importance in learned models (including both the RankSVM and the Random Forest models) into cost model heuristics. 
}

\subsection{Plan Enumeration} \label{sec:enumeration}

Given a Vega specification, VegaPlus's plan enumerator generates $n$ candidate plans for executing this specification. 
Execution plans are different ways to partition the dataflow's transform operations across the Vega runtime running in the browser and the server-side (or remote) DBMS. 


\paragraph{Data Dependency Checking} Transform operations are contained in the \texttt{data pipeline} component of the specification. The data pipeline transforms the raw data source into one or multiple smaller datasets that can be directly mapped to visual attributes such as x- and y-axes, color mappings, \etc To coordinate the computation, we first check for data dependencies to identify data results that must be preserved on the client, i.e., values explicitly referenced by the Vega runtime. Then, we enumerate candidate execution plans that satisfy the required data results and constraints. 

In the original client-side Vega specification, the existence of the intermediate results is necessary for multiple reasons: a) other specification components, e.g., \texttt{scale}, \texttt{mark}, can refer to them, b) one intermediate result can be the source for multiple children data entries to avoid duplicate computation. 
VegaPlus may not need to maintain every data entry, except for the former reason. 
Thus, as the first step of plan enumeration, we traverse the data pipeline in the specification to identify these data entries, and derive a computation graph indicating data entry dependency and data results that must be preserved. 

\paragraph{Enumerating the Data Pipeline}
Given a specification, the enumerator outputs execution plan candidates indicating where each operator needs to be executed, on the server or on the client. In theory, a specification with $n$ operators results in $2^n$ execution plans. However, in reality, there are fewer possible candidates due to the visualization dataflow search space. 

The data movement in the dataflow graph is in a single direction because the data source resides on the DBMS and the visualization view is rendered on the client at the end. Considering any path in the DAG that starts from a data source root and ends with a leaf operator (whose output is directly mapped to visual components without further transformation), there should exist one and only one \textit{split point} where the data flows from the server to the client. The split point is conceptual so that all operations that are upstream to the split point are executed on the server, and all that are downstream should be on the client. However, two paths may share the same prefix, and diverge right after an intermediate result. If the split points of both occur at the same position before the intermediate result, then we consider they share the same split point and the same server-side operations. Otherwise, the SQL queries for different paths are executed separately. 

We enumerate all possible execution plans by traversing the computation graph and permuting the way an operator is executed. 
An operator can be rewritten in SQL only if the rewriting is supported and the parent to the data entry containing the operator is not reserved by the dependency-checking process. 
Meanwhile, we make sure that, for each operator-rewritten operator, all its ancestors are rewritten and wrapped as its subqueries. 
We consolidate paths only if they have the same prefix rewritten in SQL to avoid querying redundantly. 

For the running example in \autoref{fig:rewrite}, the original bin operator is supposed to append the bin start and end values to each data item, while the aggregate operator groups and counts the items in each bin. The example candidate plan split after all data pipeline nodes, assigning all operations to be executed on the server. The extent query is set aside because it outputs a signal requested by other operators. The aggregate operator absorbs and merges with the bin operator's query when query rewriting is performed alongside the enumeration process. In such a way, the plan avoids fetching the binning result, which has the same number of rows and two more columns, compared to the raw data source. 

\subsection{A Pair-Ranking Optimizer} \label{sec:comparison-model}
VegaPlus optimizer adopts a \textit{pairwise} approach to find the best plan. Given each pair of plan candidates, the comparator model ranks them by which one is more efficient. 

\subsubsection{Plan Encoding}
To teach a model how to discern one plan from another, we need a way of encoding each plan into a vector format understood by existing modeling libraries.
To achieve this, we vectorize each enumerated execution plan $p$ to its feature vector $v$. 
A plan vector $v$ is an array of features $v = \{f_1, ..., f_k\}$ that represent the characteristics of the dataflow graph compiled from the data pipeline in an execution plan.
We extract $v$ by traversing the dataflow, maintaining a counter for all operator types and the sums of output cardinality for each operator type. Since the cardinality can span a wide range, we apply a min-max normalizer for the cardinality features in the vector. 

We omit the structural features in encoding because the operations in JavaScript/Vega are single-threaded and blocking. Within each single interaction, the operations are serialized without loops. 
The operator counts capture the distribution of operator types. A larger number of VDTs and fewer number of other client-side operations indicate a larger extent of query rewrite. 
The output cardinality of VDTs reflects both the SQL query cost and the network cost. 
These statistics may not be sufficient to derive a prediction for execution time; however, with a pairwise comparator, they are effective in distinguishing the difference between two candidates of the \textit{same} specification. 


\subsubsection{Plan Comparators}

Given a set of plan vectors for a target Vega specification, the optimizer's plan comparator must infer the best plan from the vector set.
Specifically, for each visualization query $Q$, the plan enumerator outputs candidate plans $p_1, ..., p_n$. Then, the plan comparator picks the best plan $opt(p)$ by iteratively selecting the better plan of each pair $(p_i, p_j)$ where $i < j$.
We have implemented a suite of prediction models to make these pairwise comparisons.
Specifically, given a pair of plans $(p_i, p_j)$, each comparator model predicts whether $v_i$ is preferred based on $sign(Compare(v_i, v_j))$.

%
%
We implement three types of comparator models:

\textbf{Naive model. }
The naive model is learned through a pairwise comparison of encoded plans, with binary labels indicating which is the faster plan among the two. 
\revise{We use off-the-shelf ML packages to compare two classification methods -- specifically, the linear RankSVM and Random Forest tree-based models.}

We use a RankSVM (Support Vector Machine) model~\cite{herbrich1999support} to fit the difference of encoded vectors to the target label, so that the model, after training, would assign weights to features. At inference time, the cost of a visualization plan can be calculated as the weighted sum of the features with the weights extracted from the trained model. Hence, we do not need to invoke the model $n(n-1)/2$ times for every pair to find the best plan; rather, we calculate $n$ cost values and return the one with the lowest cost. 
More formally, the model $Compare(v_i, v_j)$ is learned from a dataset where each entry $(v_i, v_j; y)$ is a plan vector pair associated with the label $y$ indicating whether $latency(v_i) < latency(v_j)$ ($y = 1$) or vice versa ($y=0$). Using the RankSVM algorithm, eventually, we train a cost model and receive a weight vector $w$ for the features, where the class of a pair is determined by the sign of $Compare(v_i, v_j) = Cost(v_i) - Cost(v_j) = w^T (v_i - v_j)$. Given a set of all vectors $V = \{v_1, ..., v_n\}$, linear regression is performed on the dataset $D = \{(v_i, v_j; y) | 1 \leq i \neq j \leq n\}$ of size $n(n-1)/2$ by minimizing the hinge loss with the optimal weights. 
The loss function is defined as: $$L = \frac{1}{|D|} \sum_{d= (v_{d1}, v_{d2}; y_d) \in D} {max(0, 1 - y_d w^T (v_{d1} - v_{d2}))}$$

\revise{
With RankSVM, we can formulate a cost function with the learned weights and return the plan with the lowest cost in linear time. The Random Forest (RF) model, on the other hand, does not generate weights for each feature via learning. In contrast, we need a wrapper around the RF model that assigns a vote for the better plan among a pair for each prediction, and then picks the best plan that receives the majority votes. 
} 


\textbf{Heuristic model. }
\revise{Since there are too many visualization scenarios for manual cost model development, we choose to generate training data and use ML models to explore the design space with the naive models. Furthermore, we explore the possibility of reverse engineering useful findings to extract rules from trained models into heuristics and cost models. }
Based on the characteristics in the plans we observe and the weights derived in the naive model, we design a few simple rules as an alternative heuristic model. The intuition here is that if the naive model is learning useful information from our training data, it should be possible to summarize what the naive model has learned using simple rules, producing a streamlined and human-interpretable ranking model. 

\revise{We implement a heuristic model with prioritized rules, where the prioritization is derived from the weights of the trained linear RankSVM model and the feature importance is calculated from RF. For example, the first rule states that a plan receives a vote if the sum of query results cardinality is smaller than the other plan by a factor of $\alpha$. If the normalized difference is within the threshold, we break the tie by the next rule that prefers plans with more aggregations on the client-side. We follow the rules to compare a pair of plans until a decision has been made or all the rules have been applied. }

Compared to the naive model, the heuristic model ranks a pair entirely based on plan characteristics without modeling their cost. We select the best plan by ranking every pair and finding the one with the most wins. Even though the heuristic model can be deployed directly without any offline training process, it costs more time to compare every pair. 

\textbf{Random model. } As a sanity check, we also implement a random model that picks a random plan in a pair. As the number of viable plans grows, we expect the random model to perform worse than the other models, providing a useful performance baseline for comparison. We use the model in Section~\ref{sec:evaluation} (Evaluation) to assess the pairwise comparison and the end-to-end performance of the naive and heuristic models. 

\subsection{Consolidating Plan Decisions Across Interactions}

When optimizing a static dashboard (i.e., without interactions), we have exactly one form of the dataflow graph to optimize, which we describe in the previous subsections.
However, in the case of interactions, the queries to be executed may change depending on how interactions introduce new Vega transforms to the dataflow graph. As a result, the best plan for one interaction may not be the best plan for other interactions, even within the same Vega specification.
To account for this variation, we collect separate plan vectors per interaction.
As interactions are triggered and the corresponding signals update, the dataflow of each execution plan re-evaluates. This re-evaluation is partial, affecting only the subset of operators that depend on the updated signal. To handle this variation, we traverse the dataflow again to extract the vectors for the operators whose timestamps are current. For an exploration session $s$ where the dashboard is updated $t$ times, we obtain a set of $t$ vectors $s_p= \{v_{p_0}, v_{p_1},...,v_{p_t}\}$, where $v_{p_0}$ is the vector for initial rendering. 

With the comparator model, each interaction episode nominates its own best choice. The chosen plans can be different based on interaction type, operators involved and interaction parameters. To consolidate the candidates and derive a final decision, we sum up the costs of all interactions for each plan candidate $p$ and return the best plan with the minimum total cost by $\arg\!\min_p \sum_{v_{p_t} \in s_p} Cost(v_{p_t})$. As for the heuristic model, we replace the cost with the number of wins for each plan.

We also allow the weights to be configurable if certain episodes are considered more significant. For example, the user might downweight the interactions that are far in the future and prefer a plan that works best for the immediate next steps. Similarly, designers might also downweight the initial rendering, since users may be more lenient for initial rendering latency (i.e., cold start) if the interactions are more seamless~\cite{battle2020structured}.

\subsection{Caching }
Similar to prior work~\cite{battle_dynamic_2016}, We follow a straightforward caching scheme with two levels: \emph{(a)} a client-side cache and \emph{(b)} server-side middleware cache. The caches store the results of executing SQL queries executed on the DBMS.
Each cache is implemented as an array of dictionary objects, where the key for each object is the SQL query string executed and the value is the result obtained for that particular query from the DBMS. Each cache has a fixed size in terms of total queries and follows the first-in-first-out replacement policy, with additional checks to avoid duplicate entries. 
When a query in the execution plan has to be executed, we check all possible sources in the order of the client’s cache, server cache and finally the DBMS for full execution. 
To avoid the cached entity being too large, we set a threshold for the size of the query result. Only if the result size is below the threshold will we store it. 

\section{Benchmark Data Generation}
Inspired by recent exploration benchmarks for DBMSs~\cite{eichmann_idebench_2020, battle_database_2020}, we contribute a benchmark to evaluate the performance of Vega and VegaPlus. To build the benchmark, we collect five real-world datasets to act as visualization data sources and scale them to different sizes as done in prior work~\cite{eichmann_idebench_2020,battle_database_2020}.
In parallel, we curate seven Vega templates covering common visualization designs, which we describe in~\autoref{sec:templates}. Finally, in ~\autoref{sec:simulation} we describe how we simulate user interactions for each template to generate interaction-driven query workloads for measuring system performance.

\subsection{Generating Visualizations From Templates} \label{sec:templates}

\begin{figure}
    \includegraphics[width=.6\linewidth]{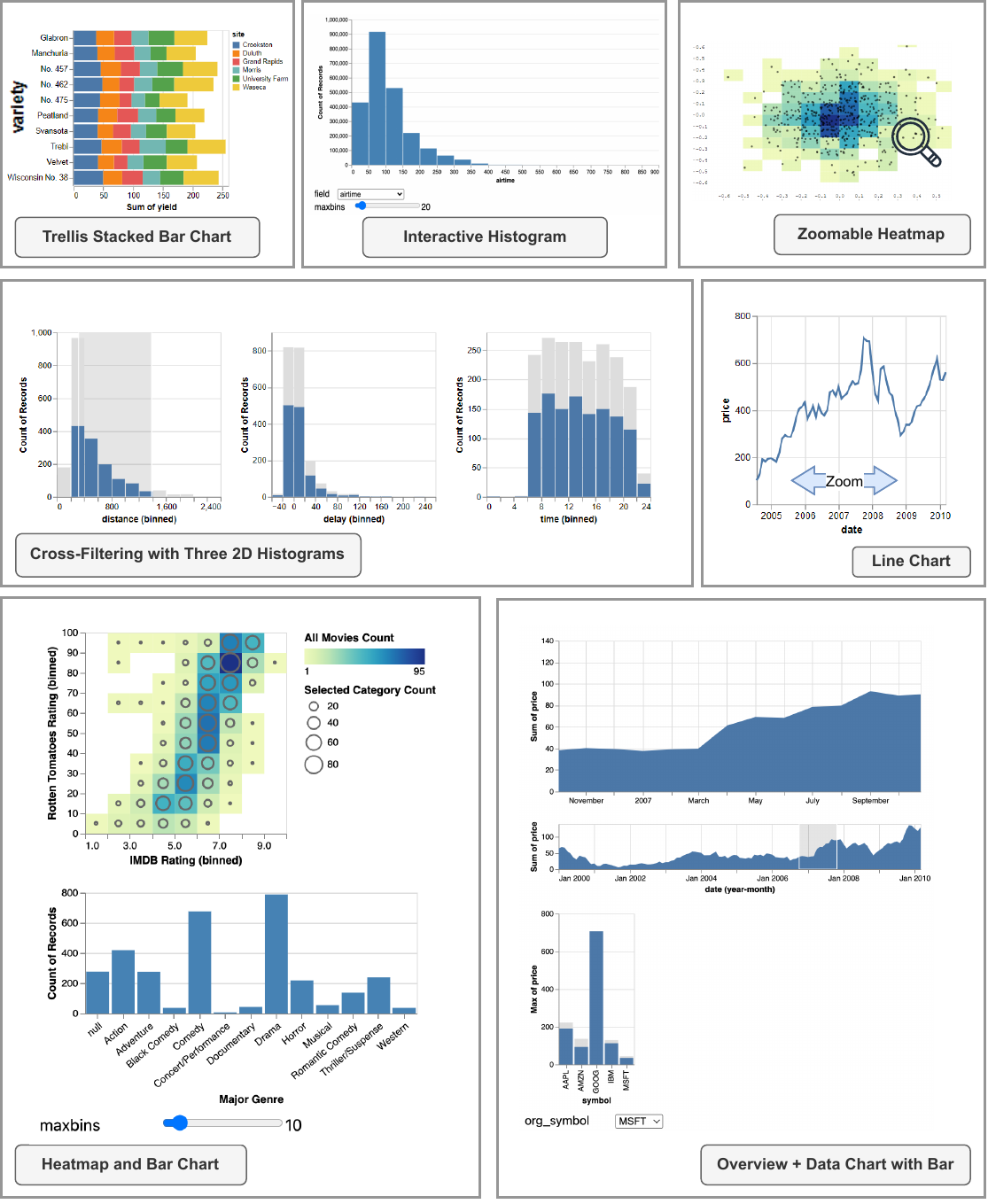}
     \caption{Templates collected for evaluation.   
     }
    \label{fig:template-overview}
\end{figure}

To generate realistic training data and assess VegaPlus's expressivity, we collect seven visualization templates including two static charts,
two single-view interactive charts,
and three interactive dashboards.
A unique feature of our benchmark is that all seven templates are independent of our source datasets, enabling us to evaluate any valid pairing of Vega code template and data source.
We display the collection in \autoref{fig:template-overview} with their native source dataset. 
Here we describe each template and explain how we parameterize them with the input datasets for training data generation. 

\textbf{Trellis Stacked Bar Chart.} This is a multi-view chart, 
where each individual view is a stacked bar chart
representing the cumulative sum of another categorical field, faceted by a third categorical field. Vega's \texttt{stack} and \texttt{aggregation} transforms are applied. 

\textbf{Line/Area Chart.} This template applies a \texttt{timeunit} transform to the \texttt{x} channel field that bins the time-series data in different intervals of choice. The encoding can be simply changed to the \texttt{area} mark, which does not affect the rest of the dataflow processing. 

\textbf{Interactive Histogram.} This template bins a quantitative field on the \texttt{x} channel and applies \texttt{aggregation} to count observed values from the \texttt{y} channel. Both the bin size and choice of the field are parameterized, allowing them to be controlled by a slider and a drop-down selection menu, respectively. 

\textbf{Zoomable Heatmap.} This chart uses 2D binning and aggregation. The panning and zooming interactions it supports trigger recalculation of binning and aggregation for density. 

\textbf{Crossfiltering With Three 2D Histograms.} This dashboard contains three histograms linked by interactions. In each histogram view, a \texttt{brush} interaction filters and re-aggregates all linked views according to the brushed filter range. 

\textbf{Heatmap and Bar Chart. } The heatmap counts observations binned along two categorical fields on the \texttt{x} and \texttt{y} channels. It is linked to a bar chart that counts the number of records for a third categorical field. By clicking on the bars in the bar chart, the heatmap filters and updates the density accordingly. An additional slider widget is provided to adjust the bin size in the heat map. 

\textbf{Overview+Detail Chart With Bar Chart. } \texttt{Interval brushes} can be applied to the overview area chart to update how data points in the detail view are binned. Then, a bar chart groups the data by a categorical field and the user can filter both the overview and detail views by clicking on individual bars.

\subsection{Parameter and Interaction Simulation}
\label{sec:simulation}
\begin{figure}
    \includegraphics[width=.6\linewidth]{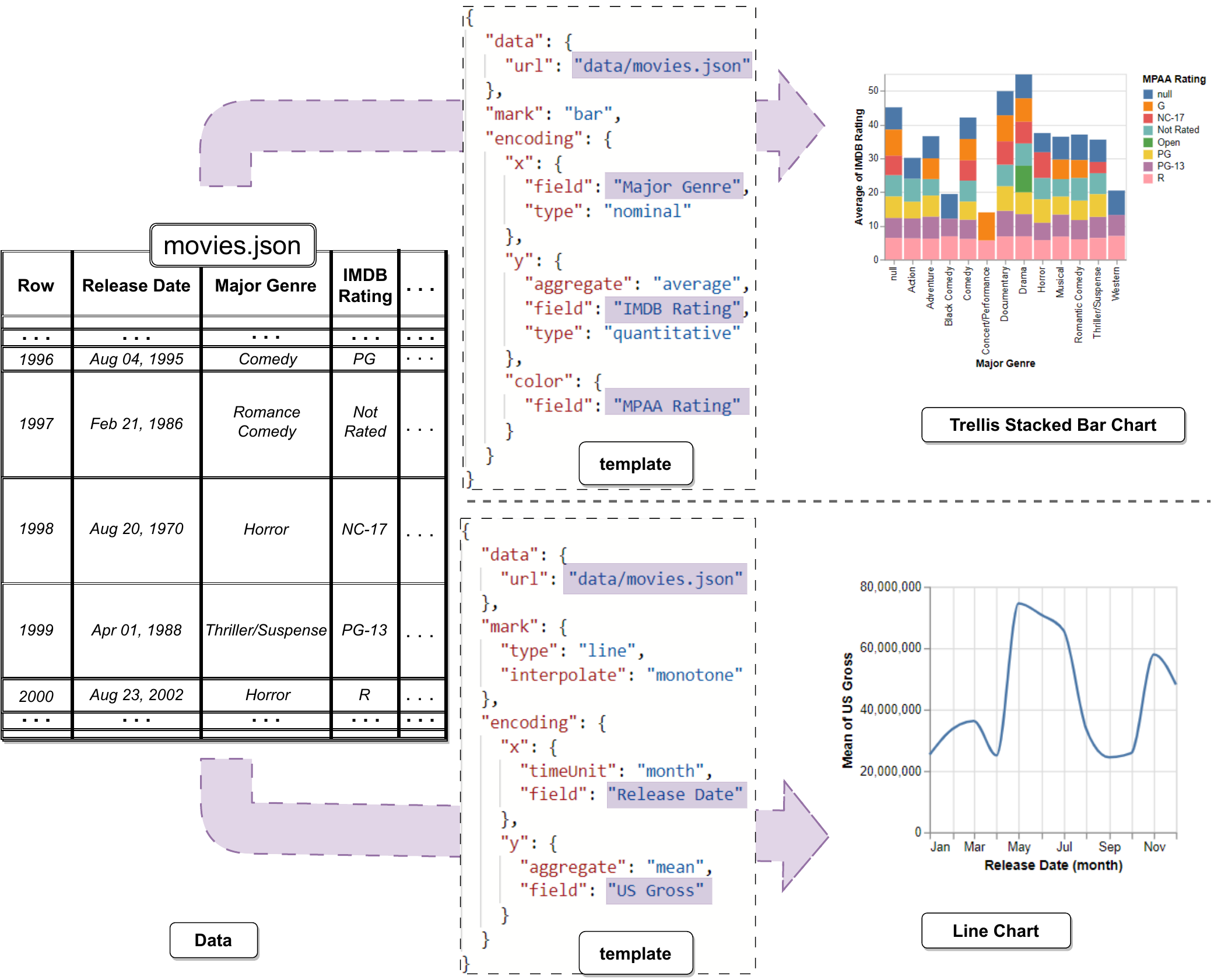}
     \caption{ Given a source dataset, specification templates are populated with values highlighted in purple. As a result, concrete visualizations can be rendered. 
     }
    \label{fig:template-spec-demo}
\end{figure}

Parameters and interactions are generated in two steps. First, we populate the template to obtain a specification for initial rendering. The process is illustrated in \autoref{fig:template-spec-demo}. Then, we simulate the interaction sequence and generate interaction parameters based on their type. 
Within each template, there are specific contents that need to be populated with data. For example, the template allows for the dynamic replacement of \texttt{data fields} in the specification. Given source data and the data type, we randomly pick from the suitable fields and fill in all the places that refer to this same field.  

To evaluate VegaPlus's performance in static and interactive contexts, we implement a workload generator to simulate potential interactions.
Here, workloads are sequences of interactions supported by the given dashboard template.
In Vega, interactions are implemented using signals. Users can specify listeners on the desired interaction/visualization components, and listeners emit data (i.e., signals) reporting the current interaction state, which can be used to update the visualization(s).
The signal types from our templates support interactions performed using separate widgets (e.g., slider filters and drop-down menu selection) as well as interactions performed directly on target visualizations (e.g., panning and zooming, brushing and linking, and selection by clicking). 

For each template, we parse the specification to collect all possible signal types, which we use to generate individual interactions. We repeatedly call this interaction generation code to produce interaction sequences to form a workload.
%
Given that data fields are selected randomly to populate our Vega visualization templates, we must gather statistics from the randomized dataset to determine the interaction/signal parameters. For example, calculating the corresponding data range for sliders and brush filters, and using this data to randomly select filter ranges for simulated interactions. To bind the input for drop-down menu selection, valid options need to be collected from the unique values in categorical data fields. 
Note that these probabilities are also configurable.



\section{Evaluation}
\label{sec:evaluation}

To evaluate VegaPlus, we first analyze the customized workload in \autoref{subsec:workload}.
Then, we examine how different models perform in the visualization initial rendering phase (\autoref{sec:initial-rendering}), and the interaction sessions (\autoref{sec:interaction}).
Finally, we compare its performance to Vega \revise{and VegaFusion as the baselines} in \autoref{sec:vega-vs-vp}.

\subsection{Experiment Setup}
\paragraph{Data}
We evaluate the VegaPlus optimizer with a customized benchmark using the templates described in \autoref{sec:templates}. For each template, we generate a workload of 10 sessions, each with 20 interactions. Given the template, each time we randomly pick one dataset as the source data and simulate the parameters and interactions. We vary the source data sizes from 50,000 rows to 1 million rows.

\paragraph{Setup} 
We deploy the VegaPlus system on a Linux machine with two 12-core 2.6 GHz Intel Xeon processors and 512GB RAM running Rocky Linux 9.1. The SQL queries are executed on PostgreSQL 15.1 with the default configuration. \revise{To compare with VegaFusion more fairly, we switch to DuckDB(v0.6.1) for both VegaPlus and VegaFusion as the SQL query execution engine. }

\subsection{Analysis of Templates and Their Plan Enumeration Space }
\label{subsec:workload}
\begin{figure*}
    \includegraphics[width=0.7\linewidth]{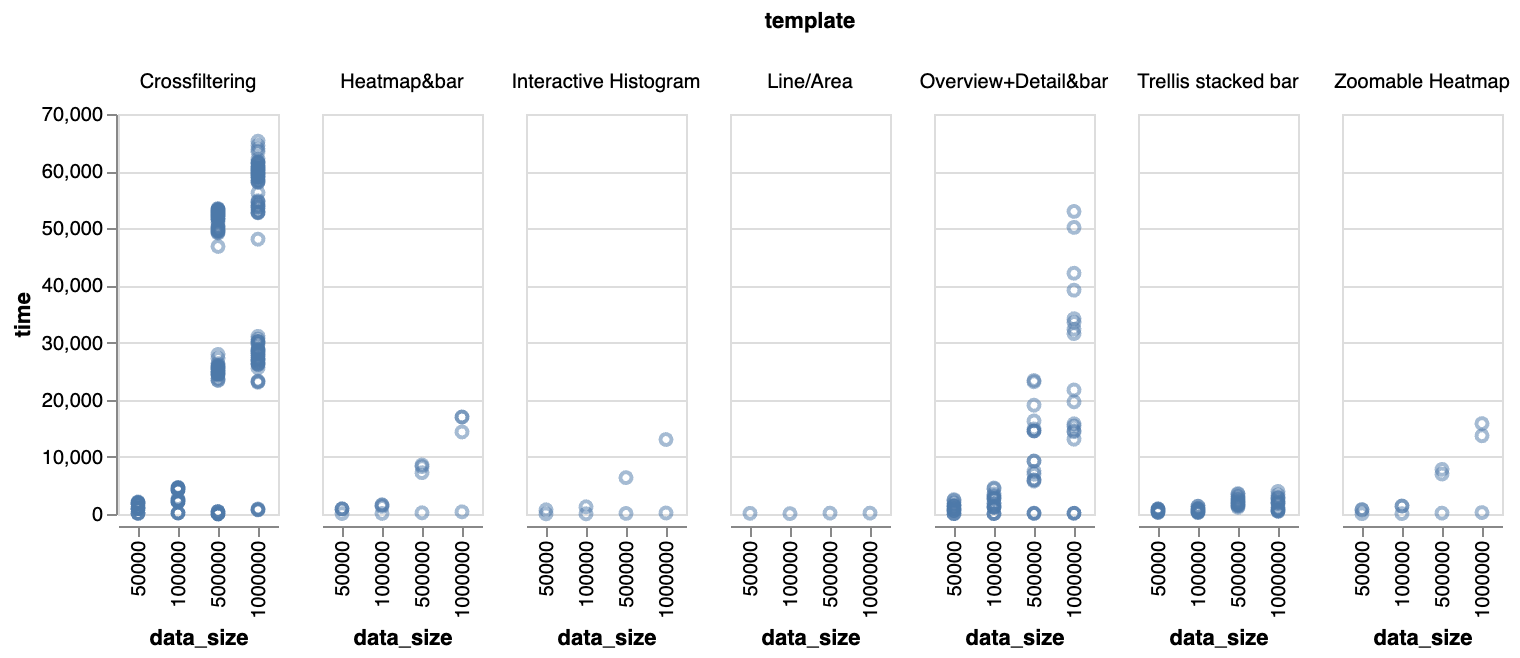}
     \caption{Distribution of plan candidates' execution time of the initial rendering for each template.
     }
    \label{fig:temp-analysis}
\end{figure*}

\paragraph{\revise{Enumeration space.}} We report the number of operators and the number of plans enumerated for each template in \autoref{table: temp-stat}.
Templates for single-view visualizations have fewer operators than the multi-view dashboards, and enumerate fewer plan candidates as a result. We obtain more training data points in pairs by generating 10 sessions each with 20 interactions for each data size from 50,000 to 1 million rows. For each template, we also report the number of data points in the same table. For a static template with $n$ candidates, we receive $10 * \binom{n}{2}$ pairs for each data size. For interactive templates, we obtain $200 * \binom{n}{2}$ pairs for each data size due to the interactions. 

\revise{
 Our template collection is representative of how it covers the most common designs and data transformation types observed in the official Vega visualization gallery~\footnote{https://vega.github.io/vega/examples/}. For most of our templates (i.e., the common case for Vega users), the combinatorial enumeration does not result in a prohibitively large training overhead. For example, training on hundreds of plans is completed within one minute, and making predictions with the random forest model takes tens of microseconds. 
 Enumerating plans for model training takes less than one second, even for the template with the most plans (Crossfiltering With Three 2D Histograms).
 That being said, pruning strategies could be applied in the future to reduce the enumeration and plan ranking time. One strategy is to use simple heuristics to
 constrain the search space. For example, we can prune all plans involving enumerated operators with an estimated output cardinality above a certain threshold. Alternatively, we can apply bottom-up boundary pruning~\cite{kaoudi2020ml-based} to reduce the search space from $O(2^n)$ to $O(n)$, where $n$ is the number of operators. }
 
\paragraph{\revise{Training data label space.}}
We execute the queries in each template (and interaction sequence) to collect the execution times, which we use to label the pairs with the most effective plan. To help understand the distribution of candidate execution time for each template, we plot the data size on the x-axis and every execution time of the 10 sessions for each plan on the y-axis as a faceted scatter plot in \autoref{fig:temp-analysis}. Note that we only plot the initial rendering without the interactions because they are more easily affected by their parameters, which introduce more factors to the analysis. We observe that templates with a larger number of candidates caused by a more complex data pipeline can have a wide range of plan execution times. Further, the plans are more sensitive to the scale of the data source: as the size increases, the plans result in higher latency. For templates such as the ``Crossfiltering with three 2D histograms'' and  ``Overview+Detail chart With Bar Chart'', there exists clusters of data points within each column of data size. And as the data size increases, the cluster margins become less distinguishable. An ideal optimizer should be able to learn these patterns. 

In general, there are significantly more slow plans than fast plans when the enumeration space is large.
A traditional pointwise model might suffer from this imbalanced plan enumeration space, given a lack of data for accurate prediction of plan execution times. However, with a pairwise model, we were able to switch from predicting execution times to the simpler problem of predicting the significance and signs of difference between plans. 
In the following subsection, we show that even when the comparator model has lower pairwise accuracy, it can output sufficiently fast plans due to its ``learning-to-rank'' nature.  

\revise{Keeping all valid candidate plans in the training data preserves the original plan space, helps us observe what information is learned by the naive model, and allows us to implement the heuristic model based on the results. However, executing all valid plans to generate labels can be time-consuming. To reduce training data collecting time as future work, we can adopt an active learning approach~\cite{Fu2012ASO} to identify a small set of representative plans to execute from which we can infer labels for non-executed plans~\cite{Ventura2021ExpandYT}. }

\begin{table}[]
\begin{tabular}{cccc}
\hline
\textbf{templates}                                                                          & \textbf{\begin{tabular}[c]{@{}c@{}}\# of \\ Operators\end{tabular}} & \textbf{\begin{tabular}[c]{@{}c@{}}\# of \\ plans\end{tabular}} & \textbf{\begin{tabular}[c]{@{}l@{}}\# of\\ pairs\end{tabular}} \\ \hline
\begin{tabular}[c]{@{}c@{}}Trellis Stacked \\ Bar Chart\end{tabular}               & 3                                                          & 4                                                      & 60                                                    \\ 
Line/Area Chart                                                                    & 2                                                          & 3                                                      & 600                                                    \\ 
Interactive Histogram                                                              & 3                                                          & 4                                                      & 1200                                                   \\ 
Zoomable Heatmap                                                                   & 5                                                          & 4                                                      & 1200                                                   \\ 
\begin{tabular}[c]{@{}c@{}}Crossfiltering With \\ Three 2D Histograms\end{tabular} & 14                                                         & 111                                                    & 1221000                                                \\ 
Heatmap and Bar Chart                                                              & 7                                                          & 5                                                      & 2000                                                   \\ 
\begin{tabular}[c]{@{}c@{}}Overview+Detail Chart \\ With Bar Chart\end{tabular}    & 8                                                          & 19                                                     & 34200                                                  \\ \hline
\end{tabular}
\caption{Characteristics of each template, and the number of plan candidates and training data records generated by the enumerator. }
\label{table: temp-stat}
\end{table}

\subsection{Optimizer Models in Comparison For Template Initial Rendering}
\label{sec:initial-rendering}
We compare the RankSVM-based model, the heuristic model and a model that randomly picks a plan in a pair, and evaluate their accuracy in predicting the better plan of each pair in ~\autoref{table:model-accuracy}. With the RankSVM model, we randomly split all collected data into a training set with 60\% of the data points and the rest 40\% for the test set.

\begin{table}[]
\begin{tabular}{lllll}
\hline
\multirow{2}{*}{\textbf{models}} & \multicolumn{4}{c}{\textbf{accuracy for input sizes}} \\
                        & 50000    & 100000    & 500000    & 1000000    \\ \hline
RankSVM                 & 0.755    & 0.781     & 0.803     & 0.744      \\ 
Random Forest           & 0.837    & 0.844     & 0.868     & 0.816    \\ 
heuristic               & 0.709    & 0.711     & 0.710     & 0.715      \\ 
random                  & 0.500    & 0.500     & 0.499     & 0.501      \\ \hline
\end{tabular}
\caption{Model prediction accuracy for plan pair comparison. }
\label{table:model-accuracy}
\end{table}

\subsubsection{Accuracy and Performance}
The \revise{Random Forest} model has the highest accuracy when trained on data with varied input sizes. 
\revise{While the RankSVM is slightly less accurate, it} does not generalize well across different input sizes. When the training and testing processes are performed on different data size, the accuracy decrease to between 0.60 and 0.65 (less than the heuristic model). The heuristic model does not require training and makes predictions based on the properties between two different plans. Thus, its accuracy relies on whether the rules capture the correct criteria of good plans as well as whether the rules are ordered to reflect how important they are. The random model has an accuracy of around 0.5 as expected, as it is predicting a winner between pairs of plans. 

\begin{table}[]
\begin{tabular}{lllll}
\hline
\multirow{2}{*}{\textbf{models}} & \multicolumn{4}{c}{\textbf{predicted execution time (ms)}}                    \\
                        & 50000           & 100000          & 500000          & 1000000         \\ \hline
RankSVM                 & 112.77          & 182.22          & \textbf{420.82}          & 890.53          \\ 
 Random Forest           & \textbf{104.61 }   & \textbf{159.24}     & 433.04    & \textbf{715.58  }    \\ 
heuristic               & \textbf{104.61}          & \textbf{159.24}          & 433.04          & \textbf{715.58}          \\ 
random                  & 1480.95         & 3602.70         & 40902.96        & 45857.96        \\ 
\textbf{optimal}  & \textbf{104.61} & \textbf{159.24} & \textbf{420.82} & \textbf{715.58} \\ \hline
\end{tabular}
\caption{Overall performance of models compared to ground truth (optimal plan).}
\label{table:model-performance}
\end{table}

 Initially, we assumed that higher accuracy in predicting between pairs of plans should translate to picking the fastest plans. However, our results suggest this is not the case.
~\autoref{table:model-performance} reports the execution time of the plan picked by each model and compares them with the actual fastest execution time collected from an exhaustive search. The heuristic model achieves the best result \revise{that is the same as the Random Forest} even though the model accuracy is slightly worse than the RankSVM-based model. These results naturally led us to question why accuracy does not seem to predict execution times, which we investigate in the next section.

\begin{figure}
    \includegraphics[width=.7\linewidth]{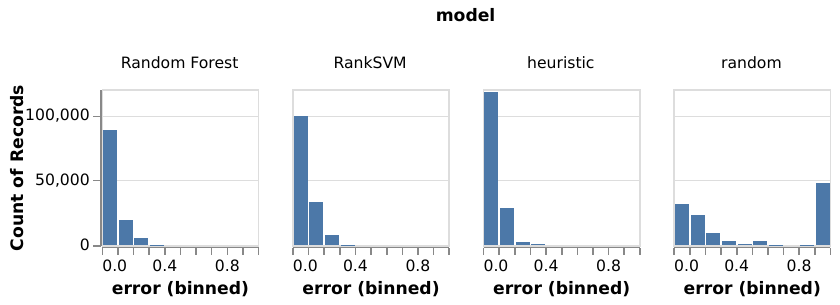}
     \caption{Distribution of scaled errors for each model.
     }
    \label{fig:model-error}
\end{figure}

\subsubsection{Distribution of Errors}
Next, we investigate how costly prediction errors were for each model. Specifically, we analyze all the incorrect predictions made by each model to see how far from optimal the prediction was in terms of execution time.
To make the results comparable across conditions, we normalize the differences in execution time to fall between 0.0 and 1.0.
We show the distribution of scaled errors for each model with input data size as 1 million rows in ~\autoref{fig:model-error}. The scaled error is measured by the average of the ratio between the difference in the execution time of the picked plan $P_i$ and the actual better plan $A_i$ with less execution time. 
\begin{equation}
    Scaled Error = \frac{1}{n} \sum_{i=1}^{n} \frac{|P_i - A_i|}{P_i}
\end{equation}
Thus,  the first bin represents the count of prediction errors for the pairs of plans with the closest execution times. Similarly, the last bin of each histogram represents the total prediction errors for the pairs of plans with the biggest difference in execution times.

First, we see that the execution time differences for all possible plans are roughly divided into clusters where the differences within are relatively small (near 0.0) and the differences between clusters are large (near 1.0). For example, the random model appears to have two error clusters in \autoref{fig:model-error} where it makes many low-cost mistakes (see the bins near 0.0) but also many high-cost mistakes (see the final 1.0 bin).
This makes sense, since the random model uses a uniform distribution to select plans, and there are significantly more bad plans than good plans in the training dataset.

While the RankSVM model is more accurate in general, it makes more incorrect predictions for pairs with larger execution time differences compared to the heuristic model.
In other words, when the RankSVM model is wrong, it tends to make costly prediction mistakes.
In contrast, the heuristic model can better distinguish the pairs and make the correct prediction when there is a larger difference in execution times.
 We see this in \autoref{fig:model-error} where the first bin of the RankSVM error histogram is lower than the first bar for the heuristic error histogram; this means the heuristic model made more mistakes with the pairs of plans with the closest execution times compared to the RankSVM model.
These findings suggest that it is not only important for a model to identify very fast plans but also to \emph{avoid very slow plans} to yield consistent, end-to-end performance gains.

\subsection{Optimizer Models in Comparison For Interactions}
\label{sec:interaction}
We compare the models' performance in interaction sessions. We generate all pairs of plans for each interaction episode as training and testing datasets. 

\paragraph{Pairwise Prediction Accuracy}
In \autoref{table:interaction-accuracy}, we report the pairwise prediction accuracy for the three models. 
The RankSVM model achieves higher accuracy compared to when it is trained only on the initial rendering data points since there is a larger number of pairs in the training dataset. On the contrary, the heuristic model with rules tailored for static dataflow generates significantly lower latency both compared to the RankSVM model and itself with only the initial rendering data points. Different from initial rendering data points, the interaction plans involve parameters of different selectivity and result in more variation for cardinality results in the encoded vectors. 
Further, the heuristic model makes decisions based on general rules such as ``prefer the plan with fewer operator X'' or ``prefer plans with a smaller sum of cardinality'', while the RankSVM model uses weights to reconcile more features. 

\paragraph{Analysis of Plan Consolidation result}
We evaluate the plan consolidation strategy and discuss the selected plan performance in terms of interaction sessions. The RankSVM model successfully selects the plans with minimum latency for all templates except for the ``Overview+Detail Chart With Bar Chart'', \revise{while the Random Forest model succeeds in this case}. However, for this template, the RankSVM optimizer picks the second-fastest plan of all, and the fastest plan is ranked third by the comparator. First, this template generates the second most number of plan candidates, following the ``Crossfiltering With
Three 2D Histograms'' template. Further, the template contains a unique \texttt{Timeunit} transform that bins the timestamp in the specified unit (\eg month), which only exists in the workflows for this template. The results are shown in \autoref{table:session-result} in comparison with the heuristic model.  
The candidates nominated by each interaction are correlated to the combination of interaction type and parameter selectivity, suggesting future work on updating the model to adapt to dynamic workflows even though our results indicate high accuracy for fixed interaction sessions. 

In \autoref{table:session-result}, we report the average performance of heuristic model selected plans and compare it with the RankSVM \revise{and Random Forest} results. The values are a summation of response time for each interaction per session, which does not include any think-time. 
We use the ``Overview+Detail Chart With Bar Chart'' template because it supports multiple interaction types and generates a relatively larger number of plan candidates.  
The heuristic model comparator consolidates decisions with the counts of nominations because the model only returns the rank of a pair without a score. As a result, the model favors the most frequent interaction types no matter how little they affect the overall session execution time. 

\begin{table}[]
\begin{tabular}{lllll}
\hline
\multirow{2}{*}{\textbf{models}} & \multicolumn{4}{c}{\textbf{accuracy for input sizes}} \\
                        & 50000    & 100000    & 500000    & 1000000    \\ \hline
RankSVM                 & 0.790    & 0.851     & 0.860     & 0.861      \\ 
Random Forest           & 0.891    & 0.887     & 0.908     & 0.891      \\ 
heuristic               & 0.621    & 0.494     & 0.576     & 0.566      \\ 
random                  & 0.499    & 0.501     & 0.499     & 0.500      \\ \hline
\end{tabular}
\caption{Model pairwise prediction accuracy with interaction episodes. }
\label{table:interaction-accuracy}
\end{table}

\begin{table}[]
\begin{tabular}{lllll}
\hline
\multirow{2}{*}{\textbf{models}} & \multicolumn{4}{c}{\textbf{predicted execution time (ms)}} \\
                        & 50000      & 100000      & 500000     & 1000000    \\ \hline
RankSVM                 & 463.26     & 584.93      & 932.70     & 1695.19    \\ 
 Random Forest           & 442.37    & 584.93     & 932.70     & 1695.19      \\ 
heuristic               & 8048.15    & 16540.91    & 90736.35   & 332776.94  \\ \hline
\end{tabular}
\caption{Average overall performance of models for template ``Overview+Detail Chart With Bar Chart'' per interaction session. }
\label{table:session-result}
\end{table}

\subsection{End-to-end Performance Versus Baselines}
\label{sec:vega-vs-vp}

\begin{figure}
    \includegraphics[width=.7\linewidth]{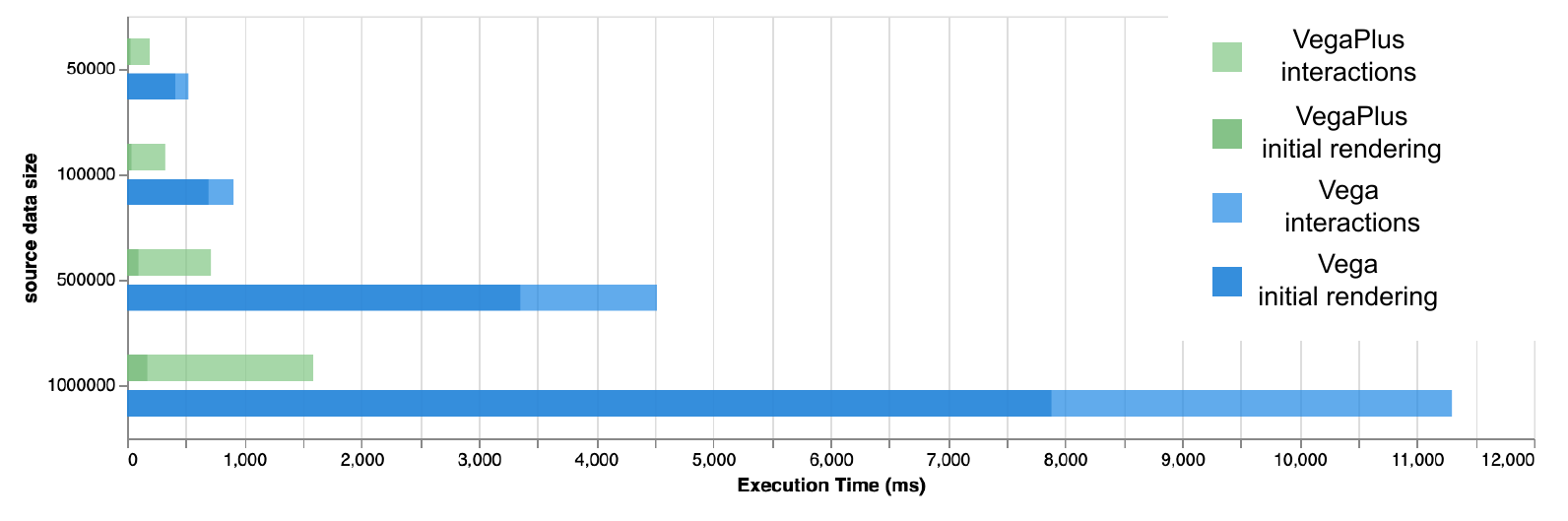}
     \caption{ Average execution time per session for Vega and VegaPlus with the RankSVM comparator model. }
    \label{fig:vega-vp}
\end{figure}

\begin{figure}
    \includegraphics[width=.7\linewidth]{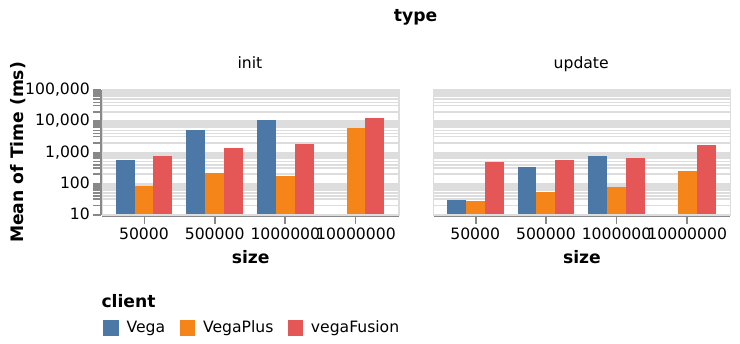}
     \caption{ \revise{Initial rendering and interactive update performance for Vega, VegaFusion, and VegaPlus (with the RankSVM comparator model). } }
    \label{fig:baselines}
\end{figure}

 We compare VegaPlus with the RankSVM comparator model against Vega, and report the results in \autoref{fig:vega-vp}. 
 Here we show the session average execution time for all interactive templates. 
 In general, VegaPlus executes specifications faster than Vega for up to 86\% with a data size of 1 million, and it scales better than Vega as the data size increases. 
 Vega takes a significant amount of time to render the initial view, which includes loading the CSV file from the disk, constructing the dataflow graph and executing the initial view. 
 The difference between Vega and VegaPlus in terms of executing iterations is smaller compared to the initial rendering. Interactions often involve a very small subset of the source data and trigger re-evaluation of a few operators in the whole data pipeline. Therefore, a large portion of interactions are not very time-consuming. 
 In fact, we observe that, for smaller data sizes (\ie 50,000 and 100,000), executing the interaction-only part takes a slightly longer time in VegaPlus than Vega. This is due to how the comparator consolidates the decision based on the session workflow. Many of the interactive templates are layered such that the data-intensive operations in initial rendering is a one-time effort. For example, the three histograms in the cross-filtering example (second row, first column in the \autoref{fig:template-overview}) require displaying the distribution of the whole source data in gray bars in the initial rendering. The operators attributed to the computation are never re-evaluated by any cross-filtering interactions. VegaPlus' RankSVM model is able to reason about the importance of episodes across a session to make the decision. It selects the candidate with faster initial rendering at the expense of a slightly slower interaction response. 

 \revise{
 We assess interactive performance with the Crossfiltering with three 2D histogram examples for Vega, VegaFusion~\cite{kruchten2022vegafusion} and VegaPlus. For VegaPlus, we choose the RankSVM comparator because it shows comparable end-to-end execution time and has a fast plan ranking time. 
 We replace the backend data processor in both VegaPlus and VegaFusion with DuckDB. 
 In \autoref{fig:baselines}, VegaPlus results in better performance for both initial rendering and interactive updates even with smaller data sizes (\ie 50,000 and 100,000) in contrast to using PostgreSQL. 
 We extend the data size to 10 million rows with VegaFusion and VegaPlus, and drop the Vega condition because it cannot handle the data size.
 Query execution time also increases with datasets; hence the underperformance in VegaFusion and VegaPlus with larger data sizes. Future work may leverage indexing techniques~\cite{moritz_falcon_2019} to reduce the latency in updating interactive linked views. 
 }
\section{Related Work}
In this section, we discuss related work in visualization languages and optimizing for data exploration in big data environments.

\subsection{Behavior-Driven Optimization}

Data scientists often explore and analyze their data in predictable patterns~\cite{battle_characterizing_2019,ottley_follow_2019,mccamish_data_2018,battle2020structured}. By modeling these patterns, DBMSs can anticipate and provision for user exploration actions through interaction-aware data caching~\cite{bavoil_vistrails_2005,callahan_vistrails_2006}, pre-fetching~\cite{battle_dynamic_2016,dimitriadou_explore-by-example_2014,sye-min_chan_maintaining_2008,singh_skimmer_2012,ottley_follow_2019,mohammed_continuous_2020,moritz_falcon_2019}, indexing~\cite{doraiswamy_gpu-based_2016,eldawy_shahed_2015,ferreira_visual_2013,psallidas_smoke_2018,tao_kyrix_2019-1,tao_kyrix_2019,zoumpatianos_indexing_2014}, and partitioning~\cite{idreos_database_2007,pavlovic_space_2016,shanbhag_robust_2017,mahmud_survey_2020} techniques.
However, these techniques are developed primarily for interactive analysis contexts where the interface design does not change, for example with a fixed dashboard design.
Certain systems enable the creation of new visualizations, but restrict the types of supported visualizations (\eg only bar charts~\cite{galakatos_revisiting_2017}) or types of supported interactions (\eg only pan/zoom interactions~\cite{tao_kyrix_2019,tao_kyrix_2019-1}).
We extend this work to support optimization in cases where the user (or the underlying system) can specify a wide range of new visualization and interaction designs.

\subsection{Optimizing Visualization Languages}

A number of projects aim to offer their users a language or API with which to design visualizations, where the data processing and rendering performance is optimized automatically by the language compiler/interpreter. Stolte~\ea~\cite{stolte_polaris_2002} present a table algebra for translating a core set of visualization designs into SQL queries and vice versa. Siddiqui~\ea present a SQL-like language for efficient searching within and rendering of a set of possible visualization designs~\cite{siddiqui_effortless_2016}. Tao~\ea~\cite{tao_kyrix_2019} present a specification language for constructing scalable pan-zoom visualizations, with a focus on scatterplot and heatmap visualizations. Ren~\ea and Li~\ea developed domain-specific languages to leverage WebGL and GPU accelerated data processing to enable fast interactive visualizations~\cite{ren_stardust_2017,li_p4_2020} and machine learning~\cite{li_p6_2020} in browsers. 
In contrast, the Vega language enables many different visualization and interaction designs~\cite{satyanarayan_reactive_2016}
but does not scale to support larger datasets~\cite{kruchten2022vegafusion}.

More recently, Krutchen et al.~\cite{kruchten2022vegafusion} extend Vega by moving certain data transformations outside of the browser to a dedicated middleware layer written in Rust.
Though effective, the range of possible visualization designs supported by these languages is limited. Furthermore, some solutions fail to match the scalability of DBMSs (e.g., \cite{kruchten2022vegafusion}).
\revise{
Quansight open-sourced the \texttt{ibis-vega-transform}~\cite{ibis-vega-transform} Python library and Jupyter extension. It supports manually composing SQL expressions with Pandas-like API (\ie ibis) to replace Vega transform pipelines and evaluating them on HeavyDB, which is optimized through hardware-accelerated parallelization. 
In this paper, we present an optimization approach that integrates the dataflow structure of Vega with new and existing database management techniques. Although VegaPlus shares the same idea of offloading operations, it automates query composing and plan selection. And it supports any user-provided backends including HeavyDB. 
}


\subsection{ML/Deep Learning for Query Optimization}

Visualization dataflows share structural similarities with DBMS query plans and therefore can potentially be optimized with similar techniques. Learning-based methods have been applied to several vital components of query optimization such as cardinality estimation~\cite{liu_cardinality_2015}, cost models~\cite{siddiqui_cost_2020} and query performance prediction~\cite{marcus_plan_2019, Venkataraman_ernest_2016}. 
Recently, ML has been used to tackle the optimization problem in an end-to-end fashion. Neo~\cite{marcus_neo_2019} uses deep learning to discover query plans
directly by predicting the best possible latency. While the performance of ML models is generally superior to that of traditional ones, the learned query optimizers still suffer from some problems such as unstable performance, high learning cost, and slow model updating~\cite{zhu2023lero}. To combat these issues, Lero~\cite{zhu2023lero} 
determines the relative order of plans rather than predicting the cost or latency to improve query optimization.
We extend these ideas to a visualization dataflow context.
Instead of learning query execution costs or latencies separately and comparing them later, 
our models directly solve the client-server partitioning problem using end-to-end plan enumeration and ranking strategies tailored to interactive data exploration environments.

\section{Conclusion}

VegaPlus is a system for automatically optimizing cross-stack visualization execution to support interactive exploration of large datasets. VegaPlus dynamically offloads computationally-intensive operations from the Vega runtime on the client side to a back-end DBMS. The interaction-aware optimizer adopts a pairwise approach to compare execution plans that partition the computation, and consolidate decisions across user interactions. We evaluate the performance and expressiveness of VegaPlus through benchmark experiments, which show improvements in scalability and interactive performance compared to the web-based Vega library. 
\section{Acknowledgments}
{The authors wish to thank colleagues in the UW Interactive Data Lab and the UW Database Lab. This work was supported in part by the NSF through award number IIS-2141506.}

\bibliographystyle{ACM-Reference-Format}
\bibliography{main}

\end{document}